\documentclass[prl,aps,twocolumn,nofootinbib]{revtex4}
\usepackage[utf8]{inputenc}
\usepackage{graphicx}
\usepackage{color}

\def\gr{$\gamma$-ray}
\newcommand{\ok}[1]{\textit{\textcolor{red}{}}}
\begin{document}

\title{Limit on intergalactic magnetic field  from ultra-high-energy cosmic ray hotspot in Perseus-Pisces region}
\author{Andrii Neronov$^{1,2}$, Dmitri Semikoz$^{1,3,4}$, Oleg Kalashev$^{3}$}
\affiliation{$^1$ Université de Paris, CNRS, Astroparticule et Cosmologie,  F-75006 Paris, France\\
$^2$Laboratory of Astrophysics, Ecole Polytechnique Federale de Lausanne, 1015, Lausanne, Switzerland \\
$^3$ Institute for Nuclear Research of the Russian Academy of Sciences, 60th October Anniversary Prospect 7a, Moscow 117312, Russia\\
$^4$National Research Nuclear University MEPHI (Moscow Engineering Physics Institute),
Kashirskoe highway 31, 115409 Moscow, Russia}

\begin{abstract}
    Telescope Array collaboration has reported an evidence for existence of a source of ultra-high-energy cosmic ray (UHECR) events in Perseus-Pisces supercluster. We show that the mere existence of such a source imposes an upper bound on the strength of intergalactic magnetic field (IGMF) in the Taurus void lying between the Perseus-Pisces supercluster and the Milky Way galaxy. This limit is at the level of $10^{-10}$~G for a field with correlation length  larger than the distance of the super-cluster ($\sim 70$~Mpc). This bound is an order-of-magnitude stronger that the previously known bound on IGMF from radio Faraday rotation measurements and it is the first upper bound on magnetic field in the voids of the Large Scale Structure. 
 \end{abstract}
\maketitle

\paragraph{Introduction.}

Telescope Array (TA) collaboration has recently reported an evidence for existence of an excess of ultra-high-energy cosmic ray (UHECR) events with energies above $E_{thr}=2.5\times 10^{19}$~eV in the direction $(RA, Dec)=(19^\circ, 35^\circ)$ of Perseus-Pisces supercluster \cite{TelescopeArray:2021dfb}.  The excess is observed at an angular scale $\Theta_{PP}=20^\circ$, similar to the angular scale of excesses  found earlier in combined HiRes/AGASA data with energies  above 40 EeV \cite{Kachelriess:2005uf}, in the data of Pierre Auger Observatory, near Cen A source \cite{PierreAuger:2010ofq} and in a previously reported "TA hotspot" \cite{TelescopeArray:2014tsd} at the energy above 57 EeV.  The new excess differs from the previously reported excesses in the sense that it has a well-identifiable counterpart, the Perseus-Pisces supercluster which is one of the largest and nearest mass concentrations in the local Universe, at the distance $D_{PP}\simeq 73$~Mpc, just behind the Taurus void of the Large Scale Structure \cite{2006MNRAS.373...45E,Boehringer:2021mix}.

UHECR  mass composition 
studies show that the fraction of heavy nuclei in UHECR flux increases  at high energies \cite{PierreAuger:2016use,PierreAuger:2017tlx}. Modelling of propagation of heavy nuclei through the intergalactic medium and through the Galactic Magnetic Field (GMF) suffer from large uncertainties that complicate interpretation of the UHECR data on previously known excesses.  The newly found TA excess \cite{TelescopeArray:2021dfb} is at lower energies and potentially can contain larger fraction of protons and helium that are easier to trace from the source to the Earth.



Trajectories of UHECR are   affected by magnetic fields in the host galaxy, galaxy cluster or supercluster, in the intergalactic medium and finally in the Milky Way galaxy. 
Strong enough magnetic field in the source located in a galaxy cluster or supercluster   may randomize the directions of UHECR particles in energy-dependent way  and change observational appearance of a source and it's spectrum \cite{Dolag:2008py}. 
The Perseus-Pisces supercluster, the suggested source of UHECR, spans a $>40^\circ$ long filament aligned with the Galactic latitude direction \cite{2006MNRAS.373...45E} in the Outer Galaxy part of  the sky. Strong magnetic field in the supercluster may well spread  UHECR across this large region and explain the extension of the TA source.

Magnetic field in the Milky Way has a different effect on the observational appearance of the source. Its regular component displaces the source from its reference position on the sky \cite{jansson12}, while its turbulent component broadens the source extent \cite{Pshirkov:2013wka}.  This broadening is by maximum $7^\circ$, moderate compared to the observed source extent \cite{Pshirkov:2013wka}. Contrary to the turbulent field, the displacement by the regular  GMF is estimated to be sizeable in the energy range of interest   \cite{jansson12},
$\Theta_{gal}\simeq 15^\circ Z \left[E/E_{thr}\right]^{-1}$
where $Z$ is the atomic number of the UHECR nuclei. It is comparable with the extent of the Perseus-Piscess supercluster on the sky and to the observed angular extent of UHECR source. 

Contrary to the GMF, presence of the intergalactic magnetic field (IGMF) between the source and the Milky Way galaxy can potentially not only transform the source appearance, but completely wash out the UHECR source from the sky. The mere existence of an isolated UHECR source thus limits the IGMF strength. In what follows,  we use this fact to derive, for the first time, an upper bound on IGMF in a void of the Large Scale Structure. 

\begin{figure*}
    \includegraphics[width=0.55\linewidth]{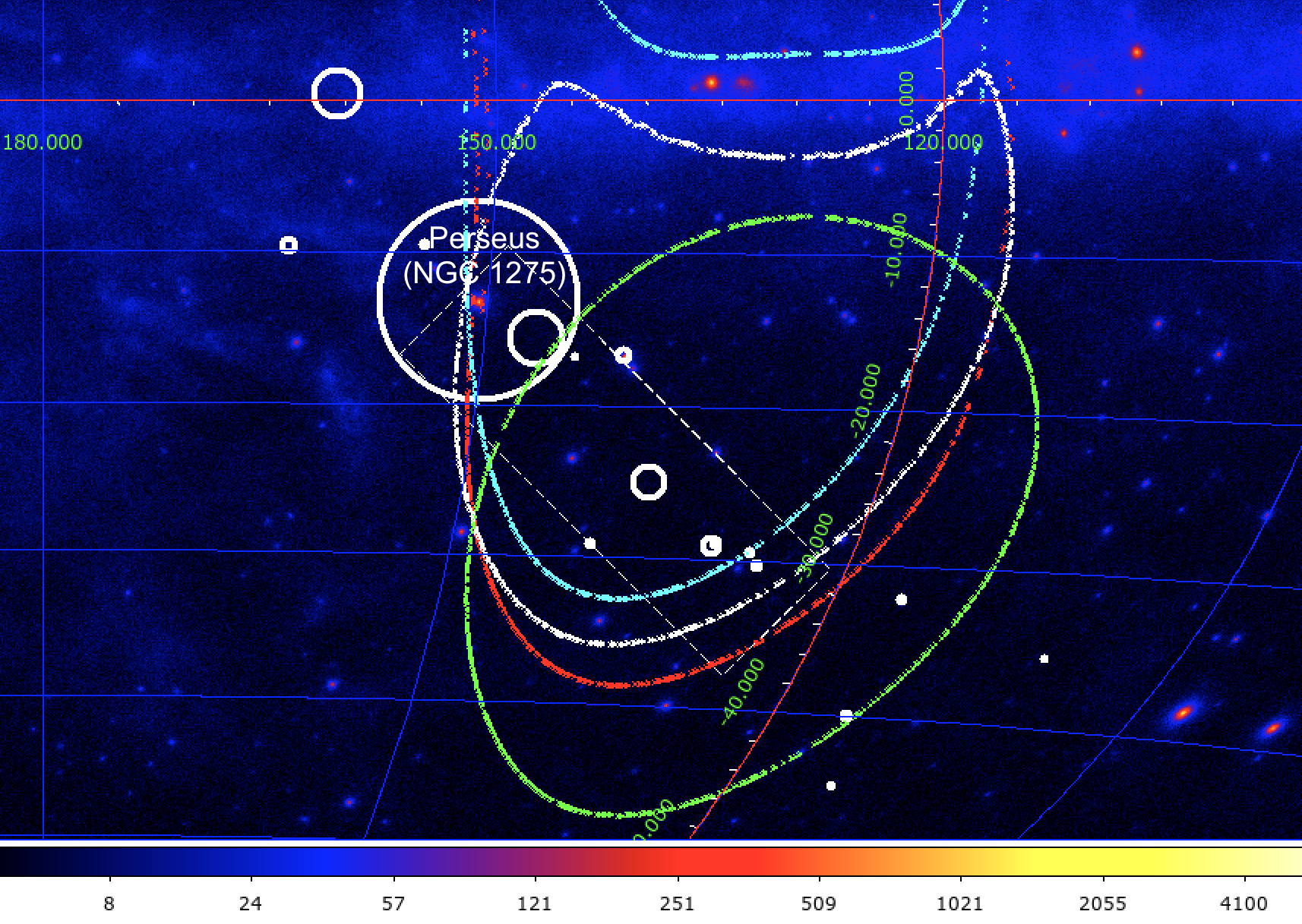}
    \includegraphics[width=0.36\linewidth]{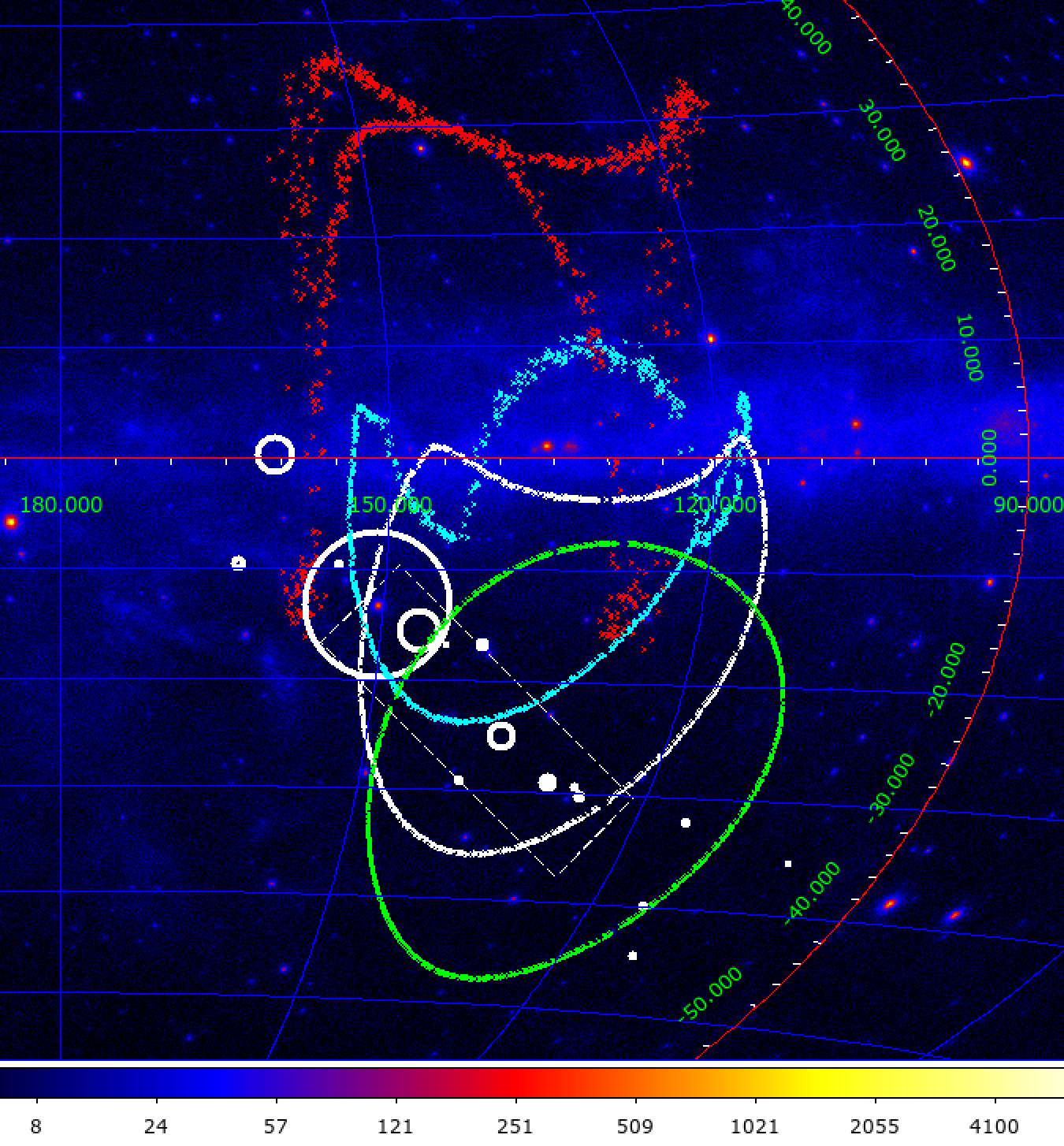}
    \caption{Left: UHECR source location chart on top of the Fermi/LAT countmap of \gr\ events with energies above 1~GeV. Green region shows the location and extent of the UHECR source of TA. Colour contours show respectively the the back-tracing of the source circular region through the GMF in different models: red for the model of Ref. \cite{jansson12}, white for the model of Ref. \cite{2011ApJ...738..192P}, cyan for that of Ref. \cite{2017A&A...600A..29T}. White circles show the locations of largest galaxy clusters and groups in Perseus-Pisces supercluster \cite{Boehringer:2021mix}. Size of the circles  is proportional to the mass of the object. The dashed rectangle of the size $30^\circ\times 10^\circ$ shows the direction of filamentary alignment of the supercluster components within the extent of the back-traced UHECR source. Right: same as the left panel, but the contours shown are the original UHECR source (green) and the back-tracing of the source through the GMF model of Ref. \cite{jansson12} assuming that UHECR are protons (white), helium (cyan) or carbon (red) nuclei. The energy of UHECR particles is $2.5\times 10^{19}$~eV. 
    }
    \label{fig:skymap}
\end{figure*}

\paragraph{Estimate of the effect of the GMF.}
Our knowledge of geometry of the GMF suffers from large uncertainties, see \cite{Jaffe:2019iuk} for a review and a  detailed discussion.
This geometry has been inferred from the data on the Faraday rotation measure of
pulsars \cite{2018ApJS..234...11H} and extragalactic sources \cite{taylor09} that provide information on the integral of projected magnetic field strength along different sky directions, weighted with the density of free electrons in the interstellar medium. Uncertainties of the distribution of  the free electrons induce large systematic errors in the modelling.
Account of the data on synchrotron emission from the interstellar medium provides information on the total strength of the field, but introduces an additional dependence of the result on the uncertain distribution of cosmic ray electrons. 

The systematic uncertainty of the GMF modelling may be estimated from comparison of different models that derive quite different overall field geometries  \cite{jansson12,2017A&A...600A..29T,Jaffe:2019iuk}. To assess this uncertainty, we compare predictions of different models for the source position on the sky. The result of back-tracing of UHECR trajectories through the regular
component of the GMF is shown in Fig. \ref{fig:skymap}. The green region shows the source location on the sky, as reported by the TA experiment (a $20^\circ$ radius circle, shown in Galactic coordinates, Aitoff projection centred at the Galactic anti-centre). Left panel of the figure shows the result of back-tracing of  the source circle through the GMF into the intergalactic space for proton UHECR. Red, white and cyan contours shows the back-traced cones for different GMF models of Refs.  \cite{2011ApJ...738..192P,jansson12,2017A&A...600A..29T}. One can see that the models agree in the general trend, deformation and shift of the circular region closer to the Galactic Plane, produced by specific direction of the ordered GMF in the outer galaxy. However, the models strongly disagree in the region close to zero Galactic latitude. The models of \cite{2011ApJ...738..192P,2017A&A...600A..29T} predict that part of the signal may be coming from the positive Galactic latitude region, not at all from the direction of the Perseus-Pisces supercluster. This large discrepancy between the model predictions is related to the larger uncertainty of the detailed structure of magnetic field at (multi)kiloparsec distances along the Galactic disk. To the contrary, the agreement between the models is better in the higher negative Galactic latitude region, where only better constrained GMF in the local interstellar medium influences UHECR. 

Turbulent component of the GMF can be taken into account for the Jansson-Farrar model \cite{jansson12,Jansson:2012rt}.
We checked that in original version of this model smearing of cosmic rays by turbulent field is below 3 degrees at 25 EeV. However  the turbulent component in Jansson-Farrar model  perhaps still an over-estimate: such strong turbulent field reduces the escape rate of low energy cosmic rays and increases the boron-to-carbon ratio of the cosmic rays flux beyond the observed value, as was shown in \cite{Giacinti:2015hva}.

\paragraph{Possible primary UHECR source(s) in Perseus-Pisces supercluster.}  
White circles in Fig. \ref{fig:skymap} show locations of the largest galaxy clusters and groups in the Perseus-Pisces supercluster. Overall, this nearby mass concentration appears filament-like on the sky \cite{2006MNRAS.373...45E} and the filament direction (outlined by the dashed box)  can be inferred from the alignment of the largest mass concentrations, visible in the figure (from \cite{Boehringer:2021mix}). The dominant mass concentration is by far the Perseus cluster. Its location is not within the extent of the UHECR source detected by the TA, but it enters the source extent once the source cone is back-traced through the GMF. This result does not depend on the choice of the GMF model.  The Perseus cluster hosts several TeV $\gamma$-ray emitting active active galactic nuclei (AGN) (NGC 1275, IC 310)  \cite{2009ApJ...699...31A,2010A&A...519L...6N}.  UHECR acceleration and interactions are expected to be associated with the TeV  \gr\ emission \cite{Neronov:2004ga}, and hence these \gr\ sources in the Perseus cluster may be the points of initial injection of UHECR in Perseus-Pisces supercluster. Improving quality of UHECR data with TAx4 experiment \cite{tax4} may be needed to verify this hypothesis and distinguish it from alternative possibilities, like injection from a large number of star-forming and star-burst galaxies distributed over the super-cluster volume. 

\paragraph{Mass composition of the UHECR source.} The right panel of Fig. \ref{fig:skymap} shows the result of back-tracing of the source circle through the GMF of Ref. \cite{jansson12} for different atomic nuclei. Increase of the atomic number $Z$ leads to larger displacement of the back-traced region toward the Galactic Plane. The back-traced region still covers part of the extent of the Perseus-Pisces super-cluster in the case of $Z=2$ (helium nuclei), but it is completely displaced to the positive Galactic latitude for $Z=6$ (carbon). This suggests that the UHECR source flux has to be dominated by protons and helium nuclei.  A source produced by a mixture of protons and helium  coming directly from the direction of Perseus cluster is generically expected to have an extent of $\gtrsim 30$ degrees at the energy $E\gtrsim 2.5\times 10^{19}$~eV, because of the energy dependence of the UHECR deflection angle.  

\begin{figure}
    \includegraphics[width=\linewidth]{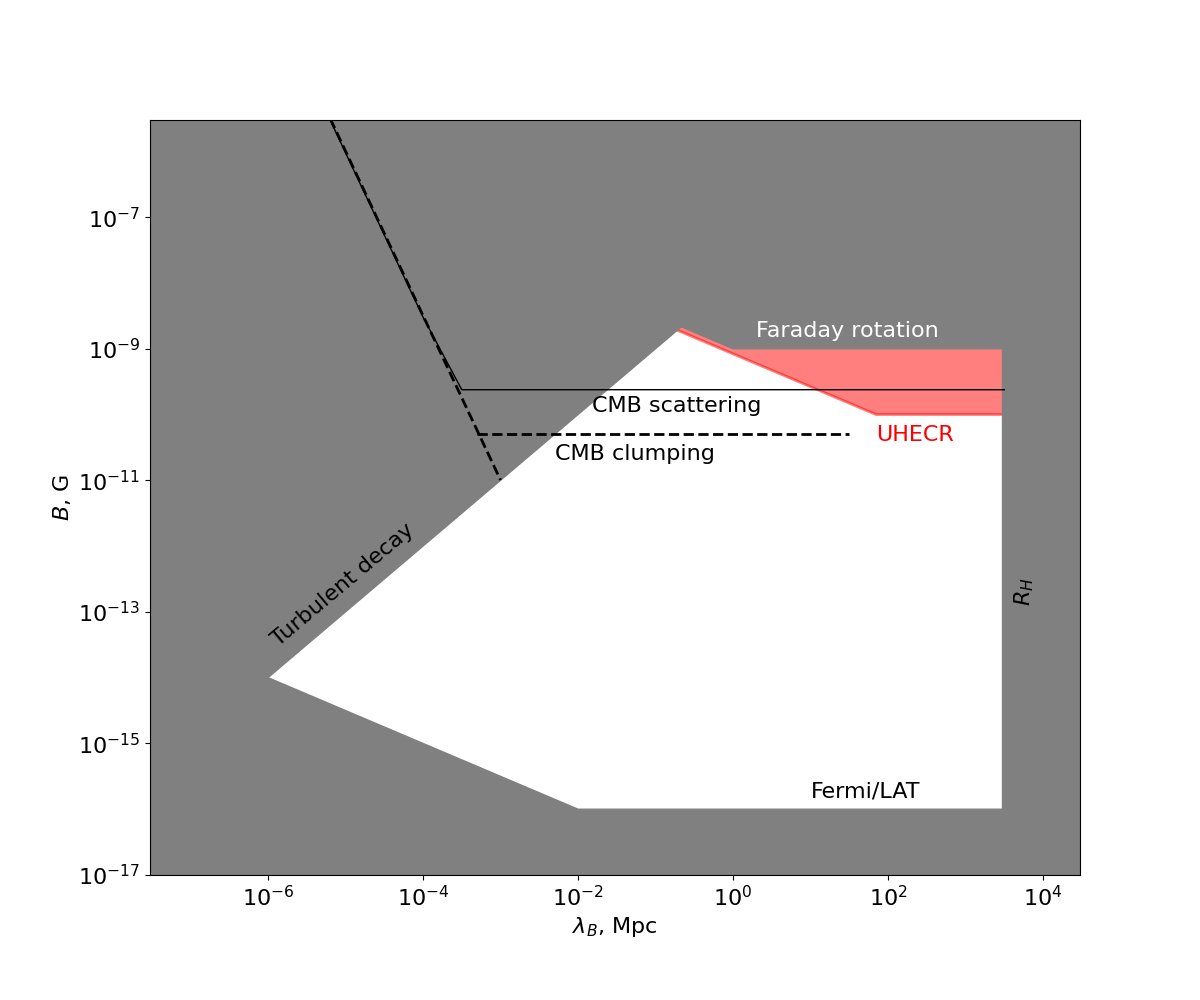}
    \caption{Known bounds on IGMF strength and correlation length (grey). Lower bound from \gr\ observations, marked "Fermi/LAT" is from \cite{neronov10,Fermi-LAT:2018jdy}. The bound marked "Turbulent decay" is corresponds to the size of largest turbulent eddies that can be processed on cosmological time scale \cite{Banerjee:2004df}. The bound marked "$R_H$" corresponds to the Hubble distance scale. The bound from Faraday rotation is from \cite{2016PhRvL.116s1302P}. The "CMB scattering" bound is from \cite{2021MNRAS.507.1254K}.  The "CMB clumping" bound is from \cite{Jedamzik:2018itu}. }
    \label{fig:exclusion}
\end{figure}

\paragraph{A limit on the intergalactic magnetic field.}
As it is mentioned in the Introduction, the mere existence of an individually detectable UHECR source (be it the Perseus cluster, or the entire filament-like Perseus-Pisces supercluster)  indicates that the IGMF is not strong enough to "erase" the source from the sky.   Homogeneous IGMF  of the strength $B$ deflects the UHECR particles by an angle \cite{neronov09}
$\Theta\simeq 15^\circ Z\left[E/E_{thr}\right]^{-1}\left[B/10^{-10}\mbox{ G}\right]\left[D/70\mbox{ Mpc}\right]$
A regular IGMF as strong  as $B\sim 10^{-9}$~G all over the distance $D\sim D_{PP}\simeq 70$~Mpc toward the source would deviate UHECR by hundred(s) of degrees so that they would even not be able to reach the Milky Way.  Thus, the source extent of $\Theta_{UHECR}=20^\circ$  imposes a limit on the homogeneous IGMF at the level of 
\begin{equation}
B\le 1.3\times 10^{-10}Z^{-1}\left[\frac{\Theta}{\Theta_{UHECR}}\right]\left[\frac{D}{D_{PP}}\right]^{-1}\left[\frac{E}{E_{thr}}\right]\mbox{ G}
\end{equation}
The uncertainty of the composition of UHECR flux does not allow to determine the characteristic atomic number of events contributing to the signal, so the conservative assumption is $Z=1$. This limit is shown as horizontal lower boundary of the red-shaded region in Fig. \ref{fig:exclusion}. 

IGMF homogeneous on the distance scale comparable to the distance to the Perseus-Pisces cluster can only originate from the epoch of Inflation in the Early Universe (see e.g. \cite{durrer13}). All other physical mechanisms  of magnetic field generation (cosmological phase transitions or magnetised outflows from galaxies) result in the fields  with much shorter correlation length: \cite{Banerjee:2004df}
$\lambda_B\simeq 1\left[B/10^{-8}\mbox{ G}\right]\mbox{ Mpc} $ 
for the phase transition field, $\lambda_B\sim 10-100$~kpc for the outflows.
In this case,  $\lambda_B\ll D_{PP}$ and the deflection angle of UHECR is estimated as \cite{neronov09}
$\Theta\simeq 15^\circ\left[E/E_{thr}\right]^{-1}\left[B/10^{-10}\mbox{ G}\right]\left[D/70\mbox{ Mpc}\right]^{1/2}\nonumber\\ \left[\lambda_B/70\mbox{ Mpc}\right]^{1/2}$.
The constraint $\Theta<\Theta_{UHECR}$ imposes a restriction $B<1.3\times 10^{-10}\left[\lambda_B/70\mbox{ Mpc}\right]^{-1/2}\mbox{ G}$.
This limit is shown as an inclined boundary of the red-shaded region in Fig. \ref{fig:exclusion}.

\paragraph{Discussion.}

Detection of an UHECR source in the direction of Perseus-Pisces cluster \cite{TelescopeArray:2021dfb} implies an upper bound on IGMF at the level shown by the red shaded range in Fig. \ref{fig:exclusion}. This bound is an order-of-magnitude tighter than the  bound from non-observation of excess Faraday rotation signal in the polarised radio fluxes from extragalactic sources \cite{1994RPPh...57..325K,2016PhRvL.116s1302P}. Moreover, this is the bound on a different kind of IGMF.  The UHECR deflections are proportional to the integral of the magnetic field component orthogonal to the line of sight: 
$\Theta_{IGMF}\propto\int_{los}B_\bot dl.$
To the contrary, the Faraday rotation measure depends on the line-of-sight integral of the parallel magnetic field weighted with the free electron density $n_e$:
$RM\propto\int_{los}B_{||} n_e dl$.
The rotation measure bound rather constrains the magnetic fields in the high $n_e$ regions, namely in halos around galaxies and galaxy clusters. To the contrary,  UHECR deflections probe average field along  the line of sight toward UHECR source.

The lines-of-sight toward the Perseus-Pisces supercluster or toward the Perseus cluster both pass through  the well-known Taurus void \cite{2006MNRAS.373...45E} (in fact, the Perseus-Pisces super-cluster is the boundary of this void, while the mass concentration hosting the Milky Way is the boundary on the other side of the void).  Thus, the limit presented above constrains the magnetic field strength in the Taurus void. This is the first time when an upper bound on the void magnetic field is reported. 

Lower bound on the magnetic field in the voids is imposed by non-observation of secondary extended and delayed  \gr\ emission around high-energy \gr\ loud AGN \cite{1995Natur.374..430P,Neronov:2006lki,neronov09}. This bound is currently at approximately $10^{-16}$~G for the large correlation length magnetic fields \cite{neronov10,Fermi-LAT:2018jdy}, see Fig. \ref{fig:exclusion}. Thus, a combination of this lower bound with the new upper bound from UHECR observations provides the first measurement of the void IGMF $10^{-16}$~G$<B<10^{-10}$~G. 
Measurements of the void IGMF will be improved in the future by the CTA telescope \cite{Vovk:2021aqb} that will be able to measure the magnetic field up to $10^{-11}$ G  \cite{Korochkin:2020pvg}.
Together with possible  improvement of UHECR observations  this can shrink the uncertainty of the void IGMF strength  to less than an order of magnitude.

The void magnetic fields are most probably of cosmological origin \cite{2001PhR...348..163G,2012SSRv..166...37W,durrer13}. However, before this can be stated with certainty, one has to verify if the voids are not "polluted"  with magnetic fields spread by galactic outflows \cite{2006MNRAS.370..319B}. 
State-of-art modelling of star formation and AGN driven magnetised outflows from galaxies \cite{illustris-tng,Garcia:2020kxm,bondarenko21} shows that these outflows most probably are not strong enough to fill the voids.  

If the void fields are of cosmological origin,  the UHECR bound on the field in the Taurus void may be compared to the bounds from the cosmological tracers of magnetic field. Presence of magnetic field during the recombination epoch induces small-scale clumping of baryonic matter and modifies the recombination dynamics \cite{Jedamzik:2018itu}. Non-observation of this effect imposes a bound on cosmological field at the level of the dashed lines in Fig. \ref{fig:exclusion}. Magnetic field also induces distortions to the matter power spectrum that affect the process of formation of dwarf galaxies \cite{2020A&A...643A..54S}. This in turn leads to earlier formation of the first stars and modification of dynamics of the reionisation. Non-observation of excess scattering of the CMB photons by free electrons generated by early reionisation  \cite{2021MNRAS.507.1254K} imposes a bound on the cosmological IGMF at the level of the solid lines in Fig. \ref{fig:exclusion}.  

The UHECR bound on large correlation length cosmological magnetic field (that would originate from Inflation)  is stronger than the limit from non-observation of magnetic field induced excess optical depth of CMB signal \cite{2021MNRAS.507.1254K} and is somewhat weaker than the limit imposed by non-observation of faster recombination \cite{Jedamzik:2018itu}. Closeness of the bounds from  three different techniques indicates that improvement of the sensitivity of these techniques may result in a "multi-messenger" detection of inflationary cosmological magnetic field, if it has scale-invariant power spectrum and the field strength close to $B\sim 10^{-10}$~G. 

The bound on IGMF derived in this paper  is comparable to previously reported constraints  from the observational evidence of correlation of UHECR arrival directions with  candidate UHECR source classes \cite{Bray_2018,VanVliet:2021sbc}, such as star-forming galaxies \cite{PierreAuger:2018qvk}. This previous analysis suffers from much larger uncertainties, compared to our analysis. The  analysis of correlation of UHECR arrival directions with entire source populations is affected by the systematic uncertainty of the deflections of UHECR by the GMF discussed above (see Fig. \ref{fig:skymap}), as well as the uncertainty of distances to the UHECR sources contributing to the correlation. The analysis of source populations also does not allow to constrain the magnetic field in the voids of the Large Scale Structure: the lines of sight to different sources forming a population may pass through voids, filaments, sheets or even highly magnetised halos.

\paragraph{Acknowledgements.} The work of D.S. and A.N. has been supported in part by the French National Research Agency (ANR) grant ANR-19-CE31-0020. Work of O.K. is supported in the framework of the State project “Science” by the Ministry of Science and Higher Education of the Russian Federation under the contract 075-15-2020-778.

\bibliography{uhecr_igmf.bib}
\end{document}